\documentclass{elsart}

\usepackage{amssymb}
\usepackage{amsmath}

\usepackage{graphicx}

\def\be{\begin{equation}}
\def\ee{\end{equation}}
\def\bea{\begin{eqnarray}}
\def\eea{\end{eqnarray}}
\def\go{\rightarrow}
\newcommand{\vF}{v_{\mathrm{F}}}
\newcommand{\kF}{k_{\mathrm{F}}}

\usepackage{feynmp}
\input hack.sty
 
\begin{document}
\setlength{\unitlength}{1truemm}

\begin{frontmatter}

\title{A model dielectric response function for metallic nanotube ropes}

\author[label1]{J. V. Alvarez}
\address[label1]{Department of Physics, University of Michigan,
         Ann Arbor, MI 48109, USA.}
\author[label2]{G. G. N. Angilella}
\address[label2]{Dipartimento di Fisica e Astronomia, Universit\`a di
   Catania\\
and Istituto Nazionale per la Fisica della Materia, UdR Catania,\\
Via S. Sofia, 64, I-95123 Catania, Italy.}
\begin{abstract}
We propose a model dielectric function for ropes of single-walled
   nanotubes distributed in a glassy graphite host medium.
We study the significance of the bosonic charge excitations arising in
   interacting quasi-one-dimensional systems in the screening
   processes.
We also pay special attention to the role of the intertube Coulomb
   interactions.
In order to compare with experiments, weak relaxation processes are
   also considered in the relaxation-time approximation.
\end{abstract}
\end{frontmatter}
\date{\today}

\section{Introduction}

Carbon nanotubes are being considered as substitutes of silicon in
   many electronic devices (see \cite{MCEUEN} for a review).
Besides their extraordinary chemical stability and  strength, they
   support ballistic electronic transport in a large range of
   temperatures.
On the basis of these properties and several experimental
   demonstrations of performance in specific devices \cite{COLLINS},
   there is an ongoing effort to find more controllable and efficient
   methods of synthesis, which are necessary in mass-production
   electronics.
Simultaneously, the importance of the one-dimensional nature of the
   electronic structure and the role of the electronic interactions in
   these materials has been recognized. The basic electronic
   properties of single-walled nanotubes (SWNT) are 
   determined by the helicity of the graphene sheet that forms the
   nanotube, \emph{i.e.} the chirality.
Nanotubes with $(n,n)$ or $(0,n)$ chiral vectors are metallic. 
In these systems the interactions, either Coulomb \cite{BALENTS} or
   phonon-assisted \cite{JISHI}, play a very important role in the
   transport properties of nanotubes at low temperatures.
The observation of Luttinger Liquid (LL) phenomenology
   \cite{EXP_LL1,EXP_LL2} has been attributed to the existence of
   strong interactions in the forward scattering channel
   \cite{KANE,EGGER}, along with the one-dimensional character of the
   electronic structure.
In addition to clear signals of non-BCS
   superconducting behavior both in ropes 
   \cite{KOCIAK} and in small-radius single nanotubes
   \cite{TANG}, other types of charge instabilities have been predicted
   theoretically \cite{yo,CARON,ALVAREZ}, deriving from the strong
   renormalization of the electron-electron couplings typical of
   quasi-one-dimensional systems at very low frequencies.      

In this paper we study the screening processes involved in a rope of
   nanotubes and we compute the dielectric response.
Our calculations are valid in the frequency range $\Delta< \omega <
   E_{c}$, where $\Delta$ is the single-particle gap associated with
   the low temperature instabilities in the rope, and $E_{c}$ is an
   energy scale, to be established later, below
   which the ropes behave as one-dimensional interacting electron
   systems. Here, we will be mainly concerned with the longitudinal response.
The real part of the conductivity can be expressed in terms of
   the imaginary part of the dielectric function via Hopfield-like
   formulas.

\section{Dielectric response}

We start the discussion giving precise meaning to the range of the
   Coulomb interaction in nanotubes.
Unlike in three dimensional metals, in strictly one-dimensional
   systems there is no conventional plasmon screening mechanism of the
   Coulomb interaction that remains long-ranged
   \cite{DasSarma,Eggert-Grabert,Belluci-Gonzalez}.
On the other hand, SWNT's have an extra dimension perpendicular to the
   main axis, namely the radius of the tube  $R$, and therefore the
   electronic wave function has an extension in this direction which
   cuts off the otherwise diverging Fourier transform $U(k)$ of the
   interaction. There are several ways to parametrize this partially
   screened Coulomb interaction. 
Here, we use one that, despite its simplicity, captures the physical
   mechanism of the screening at small momentum transfer using just
   two parameters, namely $U(k) =  \frac{\pi \vF U_{0}}{2}
   \log\left(\frac{k_c+k}{k}\right)$, and it is well motivated from
   the physical point of view \cite{Millis-DasSarma}. 
$U_0$ encodes the intensity of the interaction and $k_c$ is a
   momentum cut-off determined by a length scale associated to the
   transverse direction.

The role of the Coulomb interaction in an isolated metallic nanotube
   has been the subject of intense study, especially in relation to
   the experimental observation of Luttinger Liquid phenomenology in
   these systems.
Actually, it has been shown \cite{KANE,EGGER} that the strength of the
   backscattering and Umklapp processes mediated by the Coulomb
   potential is reduced by a relative factor $\sim 0.1(a/R)$, where
   $a$ is the lattice spacing, with respect to forward scattering
   processes.
In addition, the intertube single particle hopping is exponentially
   suppressed in typical ropes which are made of nanotubes with
   different diameters and chiralities \cite{MAAROUF}.
Under these conditions the polarization effects can be computed within
   the RPA approximation.
Let us consider then the generic non-interacting polarizability:
\be
\Pi^{(0)}=2\int \frac{dp}{2\pi}
\frac{f[\varepsilon_s (p)]-f[\varepsilon_{s'}(p+k)]}
{\omega+\varepsilon_s(p)-\varepsilon_{s'}(p+k)} ,
\ee
where $s$ and $s'$ are arbitrary band indices, and first we estimate
   the interband polarization effects.
We recall some elementary facts about the electronic band structure of
   $(n,n)$ nanotubes. For nanotubes with typical diameters 
   of $\sim 1.5$~nm, curvature
   effects are negligible and the band structure can be obtained by
   wrapping the graphene sheet and finding quantization conditions for
   the wave vector in the radial direction.
One finds
\be
  \varepsilon_s(q)=\pm
   t_{\pi}\sqrt{1+4\cos\left(\frac{\sqrt{3}}{2}qa\right)\left[\cos\left(\frac{s\pi}{L_y}\right)+\cos\left(\frac{\sqrt{3}}{2}qa\right)\right]} ,
\label{dispersion}
\ee
where $t_{\pi}\sim 2.5$~eV is the hopping integral between two carbon
   $\pi$ orbitals, $L_y$ is the perimeter of the tube measured in
   number of hexagons, and $s$ is a band index ($s=0,\pm 1,\ldots$).
Therefore, there are two bands ($s=0$) crossing the Fermi level  at
   exactly two Dirac points  $q=\pm \kF=\pm
   \frac{4\pi}{3\sqrt{3}a}$.
Linearizing the dispersion relation, Eq.~(\ref{dispersion}), we have
   $\varepsilon_{s=0} (p)=\pm \vF |p \pm \kF |$.
The next completely empty (and full) bands $s=1$ ($s=-1$) have two
   extrema at $p\sim \pm \kF$.
We can expand $\varepsilon_s (p)$ as $\varepsilon_s (p)\sim
   +E_{\mathrm{int}}+p^2/2m$, where, according to Eq.~(\ref{dispersion}),
   $E_{\mathrm{int}}=\pm t_{\pi}\sqrt{2-2\cos(\pi/L_y)}$.
For the first empty band $E_{\mathrm{int}}$ is of the order of 1.5~eV
   for nanotubes with typical diameters ($\sim 1.5$~nm).
For  momenta close to $\pm \kF$ the interband polarizability is:
\be 
\Pi^{(0)}_{\mathrm{int}}(\omega) \sim \frac{1}{\omega^2-E_{\mathrm{int}}^2}
\int \frac{dp}{2\pi}\left[f(\vF p)-f\left(\frac{(p+k)^2}{2m}
+E_{\mathrm{int}}\right)\right]
\ee
and only contributes at large frequencies.

\begin{figure}
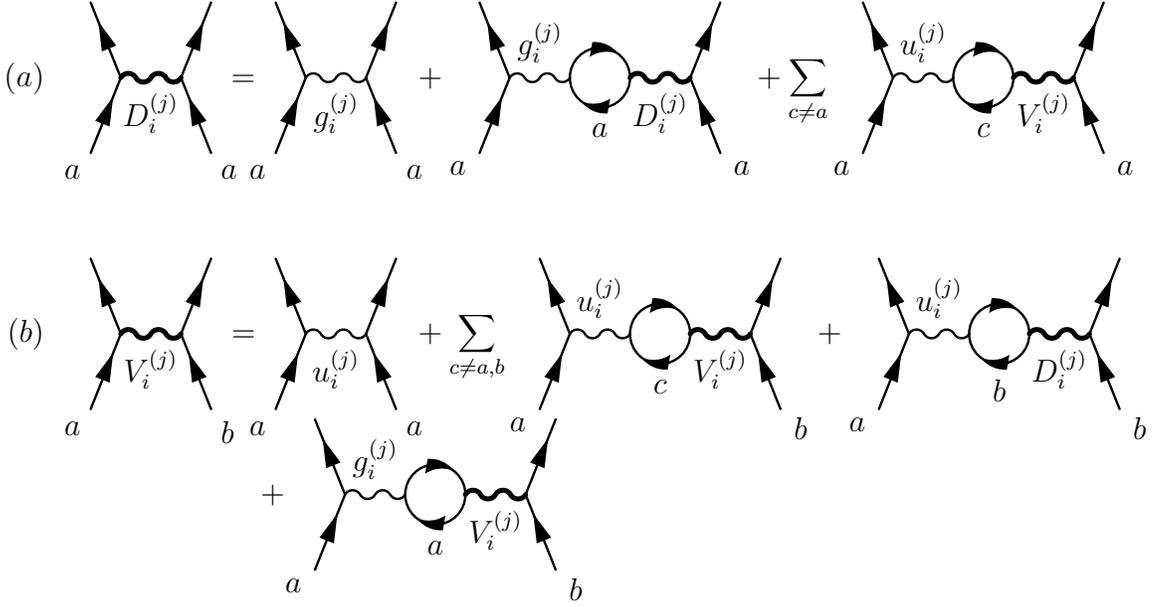

\input diagram1.tex
\caption{(a) Diagrams contributing to the screening of Coulomb
   interactions between currents $l_{a}$ within the same nanotube.
The currents have well-defined chirality and spin. 
The last term takes into account the polarization of the rest of $n-1$
   metallic nanotubes in the rope.
(b) Diagrams contributing to the screening of Coulomb interactions
   between currents $l_{a}$ and $l_{b}$ belonging to different
   metallic nanotubes. 
}
\label{rpa}
\end{figure}

On the other hand, there is a remarkable \emph{intertube} screening
   effect.
Let us call $g$ and $u$ the bare intratube and intertube coupling
   constants, respectively, and $D$ and $V$ the corresponding
   vertices.
Since the intertube polarizability between two metallic nanotubes is
   of the same order of the intratube one, the effective
   density-density interaction between two electrons located in the
   same nanotube $D_{i}^{(j)}$ is corrected by the interaction of
   electrons in different nanotubes $V_{i}^{(j)}(a,b)$.
We are using the conventional notation for super and subindices
   \cite{KROTOV}. The other pair of indices, $a$ and $b$, label different nanotubes. We consider spin independent interactions $g_{i
   \parallel}^{(j)}=g_{i \perp}^{(j)}$ and we define our bare
   couplings as $g_{i}^{(j)}=g_{i \parallel}^{(j)}+g_{i \perp}^{(j)}$
   (analogously for the $u$'s).
The virtual processes that renormalize the interactions can be
   represented in diagrammatic form (see Fig.~\ref{rpa}) and translate
   into the following Dyson equations for the vertices:
\begin{subequations}
\label{Dyson}
\begin{eqnarray}
D_{i}^{(j)}=g_{i}^{(j)}&+&\sum_{klmn}\eta_{kmi}\eta_{lnj}g_{k}^{(l)}\Pi D_{m}^{(n)}  \nonumber \\
&+&\sum_{klmn} \sum_{c \neq a} \eta_{kmi}\eta_{lnj}u_{k}^{(l)}(a,c)\Pi V_{m}^{(n)}(c,a) \label{intra} \\
V_{i}^{(j)}(a,b)=u_{i}^{(j)}(a,b)&+&\sum_{klmn}\sum_{c \neq a,b} \eta_{kmi}\eta_{lnj}u_{k}^{(l)}(a,c)\Pi V_{m}^{(n)}(c,b)  \nonumber \\
&+&\sum_{klmn}\eta_{kmi}\eta_{lnj}\Pi [u_{k}^{(l)}(a,b) D_{m}^{(n)}  +g_{k}^{(l)}V_{m}^{(n)}(a,b)] \label{inter} 
\end{eqnarray}
\end{subequations}
where $\eta_{kmi}=\delta_{4i}\delta_{km}+\delta_{2i}(1-\delta_{km}); ~~ 
\eta_{lnj}=\delta_{4j} \delta_{ln}+\delta_{2i}(1-\delta_{ln})$  
and $\Pi^{(0)}=\delta_{jl}\Pi^{(0)}_{+}+(1-\delta_{jl})\Pi^{(0)}_{-}$ ,
$\delta_{j'l'}$ are Kronecker $\delta$ functions, and $\Pi^{(0)}_{+}$
and $\Pi^{(0)}_{-}$ are the respective polarizations for right and
left branches with linear dispersion, given by
\begin{equation}
\Pi^{(0)}_{+}(k,\omega)=\frac{ k}{2\pi(\omega- \vF k)};~~~~\Pi^{(0)}_{-}(k,\omega)=\frac{-k}{2\pi(\omega+\vF k)}.
\label{pol}
\end{equation}
superscript zero refers to the pure case.

For a single nanotube we can decouple the equations in the form
\begin{subequations}
\label{eq:Dpm}
\bea
D_{\pm}^{(4)} &=& 
g_{\pm}^{(4)}+g_{\pm}^{(4)}\Pi^{(0)}_{+}D_{\pm}^{(4)}
+g_{\pm}^{(2)}\Pi^{(0)}_{-}D_{\pm}^{(2)} \\
D_{\pm}^{(2)} &=&
   g_{\pm}^{(2)}+g_{\pm}^{(2)}\Pi^{(0)}_{+}D_{\pm}^{(4)}
+g_{\pm}^{(4)}\Pi^{(0)}_{-}D_{\pm}^{(2)}  .
\eea  
\end{subequations}
where we used the combinations ${}_{4}^{(4)}\pm {}_{2}^{(4)}={}_{\pm}^{(4)}$
and ${}_{4}^{(2)}\pm {}_{2}^{(2)}={}_{\pm}^{(2)}$.
Using the Eqs.~(\ref{pol}) we obtain the expressions for the vertices.  
For instance, $D_{4}^{(4)}$ in the positive branch is: 
\be
D_{4}^{(4)}(k,\omega)=(\omega-\vF k)
\left[
  \frac{C_{+}^{+}}{\omega+u_{+}k}
+ \frac{C_{+}^{-}}{\omega-u_{+}k} 
+ \frac{C_{-}^{+}}{\omega+u_{-}k}
+ \frac{C_{-}^{-}}{\omega-u_{-}k}
\right]
\label{D44_pos}
\ee
where 
\begin{equation}
u_{\pm}=\sqrt{\left[\vF +\frac{1}{2\pi}(g_{4}^{(4)}\pm
   g_{2}^{(4)})\right]^2
-\left[\frac{1}{2\pi}(g_{4}^{(2)}\pm g_{2}^{(2)})\right]^2}
\label{velocities}
\end{equation}
and
\begin{subequations}
\begin{eqnarray}
C_{+}^{\pm}&=&\frac{1}{4}\left(g_{4}^{(4)}+g_{2}^{(4)}\right)
\pm \frac{\pi}{2u_{+}}
\left(\vF^{2} - u_{+}^{2} + 
\frac{\vF}{2\pi}(g_{4}^{(4)}+g_{2}^{(4)})\right)\\
C_{-}^{\pm}&=&\frac{1}{4}\left(g_{4}^{(4)}-g_{2}^{(4)}\right)
\pm \frac{\pi}{2u_{-}}
\left(\vF^{2} - u_{-}^{2} + 
\frac{\vF}{2\pi}(g_{4}^{(4)}-g_{2}^{(4)})\right) .
\end{eqnarray}
\end{subequations}

In the limit of frequencies larger than the ``plasma'' and free
   electron energies ($\omega \gg u_{\pm}k,\vF k$) we
   recover the value of the bare coupling $D_{i}^{(j)}(\omega
   \rightarrow \infty,k)=g_{i}^{(j)}$. 
If all the bare intratube couplings $g_{i}^{(j)}=U(k)$, the intratube
   vertex Eq.~(\ref{D44_pos}) takes the form
\begin{equation}
D_{4}^{(4)}(k,\omega)=U(k)
\frac{\omega^2 - (\vF k)^2}{\omega^2 - (u_+ k)^2} ,
\end{equation}
from which one can extract the dielectric function as ($D_{4}^{(4)}
   \equiv U / \epsilon^{(0)}$):
\begin{equation}
\epsilon^{(0)} (k,\omega) =1-\frac{2U(k)}{\pi\vF}\frac{(\vF k)^2}{\omega^2- (\vF k)^2} .
\label{eq:eps0}
\end{equation}
This result can be obtained by using the alternative route of finding
the elementary bosonic excitations using the bosonization
technique, and after that computing the density-density
correlations.

Let us now consider the problem of a bundle of $n$ nanotubes.
We assume that the perpendicular size of the rope is small and
   therefore, in first approximation, electrons in  different
   nanotubes interact with the same bare strength
   $u_{i}^{(j)}(a,b)=u_0$, whereas $g_{i}^{(j)}=g_0$.
Then Eqs.~(\ref{Dyson}) for the vertices can be rearranged in compact
   matrix form as
$\boldsymbol{\epsilon}^{(0)} \left( D \,\, V \right)^\top =
   \left( g_0 \,\, u_0 \right)^\top$,
where the $2\times 2$ matrix
\begin{equation}
\boldsymbol{\epsilon}^{(0)} = \begin{pmatrix} 
1-2g_0 \Pi^{(0)} & -2(n-1)u_0 \Pi^{(0)} \\
-2u_0 \Pi^{(0)} & 1 -2[g_0 + (n-2)u_0 ] \Pi^{(0)}
\label{eq:epsmatrix}
\end{pmatrix}
\end{equation}
generalizes the dielectric function to the case of intratube and
   intertube screening, and $\Pi^{(0)} = \Pi^{(0)}_+ + \Pi^{(0)}_-$.

The two eigenvalues of $\boldsymbol{\epsilon}^{(0)}$,
   Eq.~(\ref{eq:epsmatrix}), give a measure of the screening due to
   both intratube and intertube scattering. We  assume the
   same functional dependence for both intratube
   and intertube couplings, $g_0 = U_1 (k)$ and $u_0 = U_2 (k)$,
   both given by $U(k)$ but now characterized by different
   intensities, $U_{01}$ and $U_{02}$, and different cutoff momenta,
   $k_{c1}$ and $k_{c2}$, respectively.
In particular, different cutoff momenta can be related to the
   different radii of a single nanotube and of a nanotube rope. 
In that case, the eigenvalues of Eq.~(\ref{eq:epsmatrix}) can
   still be expressed analytically in terms of the total polarizability
   $\Pi^{(0)}$, and will be discussed numerically in the presence of
   impurities and of a host medium 
   below.

\begin{figure}[t]
\centering
\includegraphics[height=0.9\columnwidth,angle=-90]{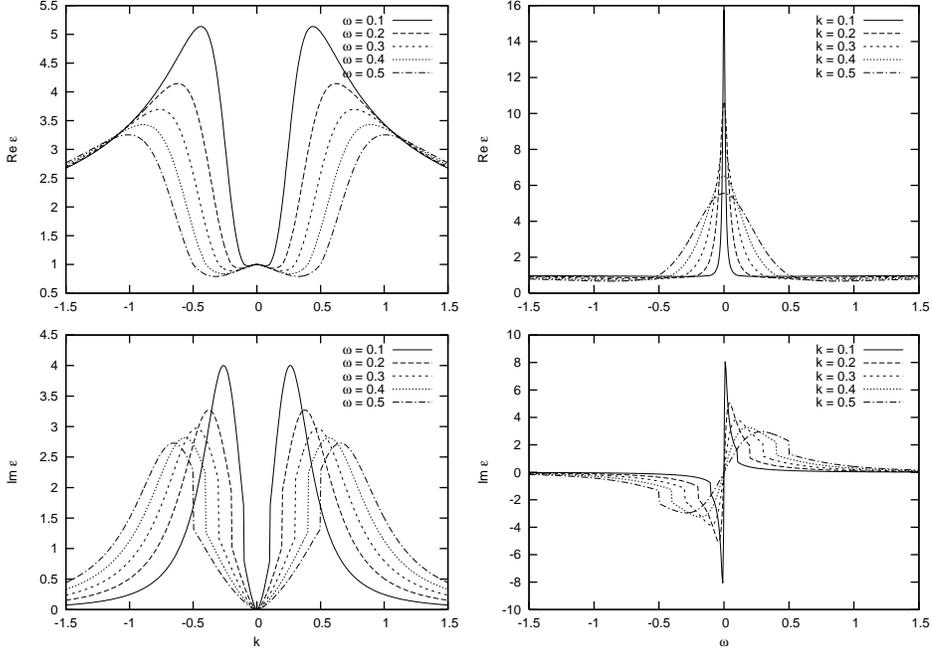}
\caption{Momentum (left panels) and frequency (right panels)
   dependence of the real (upper panels) and imaginary (lower panels)
   parts of the larger dielectric eigenvalue of
   Eq.~(\protect\ref{eq:epsmatrix}), including weak relaxation
   effects. 
Momentum  and frequency are rescaled as $\vF \tau k$ and $\omega\tau$,
   respectively.
We take $n=10$, with $U_{01} = 7$, $U_{02} = 2$, $\vF \tau k_{c1} =
   0.25$, and $\vF\tau k_{c2} = 0.05$.
}
\label{fig:epsp1}
\end{figure}

The \emph{low
   energy sector} of the theory is integrable and therefore we 
   expect  finite weight for the
   Drude delta peak in the conductivity at zero frequency.
In order to make contact with experiments, we need to introduce weak
   relaxation effects, due to collisions with defects or impurities.
We can introduce these effects in the non-interacting polarizability
   within the relaxation time approximation, which is equivalent to
   carry out the substitution \cite{Mermin,LI}:
\begin{equation}
\Pi^{(0)} (k,\omega) \go  \frac{(1+i/\omega\tau)
\Pi^{(0)} (k,\omega+i/\tau)}
{1+(i/\omega \tau) \Pi^{(0)} (k,\omega+i/\tau)/\Pi^{(0)} (k,0)} ,
\label{subst-Pi}
\end{equation}
to the \emph{total} polarizability $\Pi^{(0)} = \Pi^{(0)}_+ +
   \Pi^{(0)}_-$.
Here, $\tau$ is an \emph{external} relaxation time which can be extracted
   phenomenologically by comparison and fitting of experimental
   results \cite{saito,Bommeli1}.
In the case of a single nanotube, Eq.~(\ref{eq:eps0}), or for a rope of $n$
nanotubes with intertube interaction simply proportional to the 
intertube one $U_{\rm int}(k)=\eta U(k)$, we explicitly find
\begin{equation}
\epsilon(k,\omega) = 1 - \tilde{U}_0 \log \left( \frac{k_c + k}{k}
   \right)
\frac{k^2}
{\omega^2 -k^2 +i\omega} ,
\label{eq:epswk}
\end{equation}
where momenta are measured in units of $(\vF\tau)^{-1}$ and
   frequencies in units of $\tau^{-1}$, and $\tilde{U}_0 =
   [1+\eta(n-1)]U_0$. In the case of a rope of $n$ nanotubes, 
   with intratube and intertube
   interactions parametrized by $U_1 (k)$ and $U_2 (k)$, respectively,
   one has to perform Mermin's substitution for the total
   polarizability, Eq.~(\ref{subst-Pi}), in the elements of the dielectric
   matrix, Eq.~(\ref{eq:epsmatrix}), \emph{before} solving the
   eigenvalue equation.
Fig.~\ref{fig:epsp1} shows the real and imaginary parts
   of the larger dielectric eigenvalue, as a function of $k$
   (left panels), and of $\omega$ (right panels).

Our results are consistent with those found in semiconductor
   quantum-wire nanostructures \cite{DasSarma,QW}.
Specifically, a system of two arbitrarily close wires with equal
   densities and effective masses is formally equivalent to the single
   nanotube we have considered.
In our case, $g_i^{(j)}=U(k)$ and the acoustic plasmon is degenerated
   with the continuum.
However, strong renormalization effects are expected at low
   frequencies \cite{yo,CARON,ALVAREZ} and it would be interesting
   (but beyond the scope of this communication) to study the impact of
   such effect in the plasmon peaks, as observed in far infrared
   spectroscopy or Raman scattering.
Likewise, small-radius nanotubes seem good candidates for the
   observation of three plasmons and an analysis of plasmon stability
   and spectral weight.
Following Ref.~\cite{QW}, one can define the dispersion relations
   associated with the RPA collective modes as the zeroes of
   $\det\boldsymbol{\epsilon}$ in Eq.~\eqref{eq:epsmatrix}, now taking
   into account also the effects of impurities.
One finds two plasmon-like modes, linearly dispersing in the
   long-wavelength limit.
Specifically, the lower-frequency mode becomes more
   damped with increasing number $n$ of nanotubes in a rope, which is
   consistent with the results of Ref.~\cite{QW}, where the lower
   (acoustic) mode approached the Landau damping region with decreasing
   spatial separation between the two wires.

\begin{figure}[t]
\centering
\includegraphics[height=\columnwidth,angle=-90]{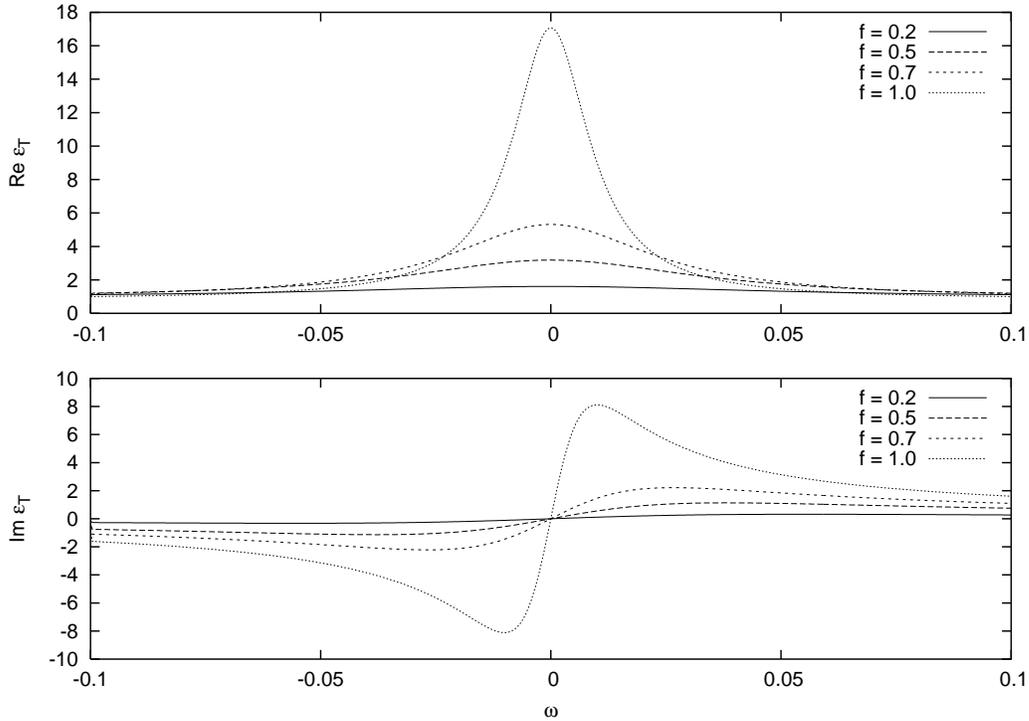}
\caption{Real (upper panel) and imaginary parts (lower panel) of total
   dielectric response, Eq.~(\protect\ref{eq:epsT}), corresponding to
   the larger eigenvalue of the dielectric matrix,
   Eq.~(\protect\ref{eq:epsmatrix}), as a function of
   $\omega\tau$, for $\vF\tau k=0.1$, $\epsilon_h =
   1$, and different values of the filling factor $f$.
All other parameters are as in Fig.~\protect\ref{fig:epsp1}.
}
\label{fig:epsT}
\end{figure}

So far, we have considered the intrinsic properties of a nanotube
   rope.
In realistic samples the nanotubes are scattered in a medium of glassy
   graphite produced during the synthesis.
Following \cite{Bommeli1} we will now assume that the
   nanotubes are placed in a host medium of insulating glassy graphite
   with a dielectric constant $\epsilon_h$.
The dielectric reponse of such a composite is given by a generalized
   Maxwell-Garnett expression \cite{COHEN}
\be
\frac{\epsilon_T}{\epsilon_h}=
\frac{2(1-f)+(1+2f) (\epsilon/\epsilon_h )}
{(2+f)+(1-f) (\epsilon/\epsilon_h )} ,
\label{eq:epsT}
\ee
where $f$ is the filling fraction of the nanotubes. 
Expression~(\ref{eq:epsT}) interpolates between the the host
   dielectric function at $f=0$ and the intrinsic value
   $\epsilon(\omega)$, given by Eq.~(\ref{eq:epswk}), at $f=1$.
The value of the host medium $\epsilon_h$ can be computed within a
   classical dispersion theory \cite{Bommeli1}.
We assume that $\epsilon_h$ is nearly constant in the frequency range
   we are interested.
The most valuable information that can be extracted when comparing
   with real experiments is the position and width of the maximum in
   the conductivity  at low frequencies (which is simply related to
   the dielectric constant). In Fig.~\ref{fig:epsT} we show 
  the total dielectric response in a medium
   with $\epsilon_h = 1$ as a function of $\omega\tau$ for different
   values of the filling fraction $f$.

In conclusion, we have studied the nature of the screening processes
   of metallic single-walled nanotube ropes.
We have considered different factors contributing to the longitudinal
   dielectric response of these systems: intratube and intertube
   Coulomb interactions, the presence of a glassy graphite environment
   and the influence of weak relaxation effects produced by impurities
   or defects. Our results suggest that metallic nanotubes may be 
   a new playground for the study of collective 
  charge excitations    

\begin{ack}
The authors thank J. W. Allen, P. Ballone, I. A. Howard, R. O. Jones,
   G.~Piccitto, R. Pucci, A. A. Varlamov for stimulating discussions. 
\end{ack}

\end{document}